\documentclass{jps-cp}

\usepackage{txfonts} %Please comment out this line unless the txfonts package is availabe in your LaTeX system.
\graphicspath{{../Doctor_thesis/epsfile/}}

\title{Importance of the Volume Fluctuation Correction on Higher Order Cumulants}

\author{Tetsuro \textsc{Sugiura}$^{1}$}

\inst{$^{1}$University of Tsukuba, 1-1-1, Tennodai, Tsukuba, Ibaraki 305-8577, Japan}

\email{tsugiura@rcf.rhic.bnl.gov }

\recdate{January 18, 2019}

\abst{   
Initial volume fluctuation (VF) caused by participant fluctuation would be the
background which should be subtracted experimentally from measured higher-order cumulants. 
STAR experiment has been applying Centrality Bin Width Correction (CBWC) to suppress VF.
However, there might be some residual fractions of VF backgrounds even with CBWC.
Recently, Volume Fluctuation Correction (VFC) has been developed under the assumption of the independent particle production (IPP) model.
In this talk, the importance of subtracting VF and validity of the VFC are studied by using simple toy models assuming IPP as well as UrQMD model.
The results showes that VFC works well in toy model but does not work well in UrQMD, which imply that IPP model could be broken in UrQMD.
}
\kword{Quark Gluon Plasma, Event-by-event fluctuation, volume fluctuation, net-charge}

\begin{document}
\maketitle

\section{Introduction}
In higher order event-by-event fluctuation analysis, initial volume fluctuation (VF) is one of the experimental backgrounds which
should be taken into account.
In order to remove VF, STAR experiment has been applying Centrality Bin Width Correction (CBWC) \cite{Xiao2} .
In CBWC, cumulants for each centrality bin are calculated by taking weighted average for each multiplicity bin as follows:
\begin{eqnarray}
C_{n}=\sum_{r}{w_{r}C_{(n,r)}},~~~~
w_{r}=\frac{N_{r}}{\sum_{r}N_{r}},
\end{eqnarray}
where $N_{r}$ and $C_{(n,r)}$ are number of events and $n^{th}$-order cumulants in $r^{th}$ multiplicity bins respectively.
Recently, a new correction method called Volume Fluctuation Correction (VFC) \cite{VFC} is proposed.
Up to the fourth-order cumulnats can be written as
\begin{eqnarray}
\label{eq_vfc1}
\kappa_{1}{(\Delta{N})}&=& \langle{N_{W}}\rangle\kappa_{1}{(\Delta{n})},\\
\label{eq_vfc2}
\kappa_{2}{(\Delta{N})}&=&\langle{N_{W}}\rangle\kappa_{2}{(\Delta{n})}+\langle{\Delta{n}}\rangle^2\kappa_{2}{(N_{W})},\\
\label{eq_vfc3}
\kappa_{3}{(\Delta{N})}&=&\langle{N_{W}}\rangle\kappa_{3}{(\Delta{n})}+3\langle\Delta{n}\rangle\kappa_{2}{(\Delta{n})}\kappa_{2}{(N_{W})} +\langle{\Delta{n}}\rangle^3\kappa_{3}{(N_{W})},\\
\label{eq_vfc4}
\kappa_{4}{(\Delta{N})}&=&\langle{N_{W}}\rangle\kappa_{4}{(\Delta{n})}+4\langle\Delta{n}\rangle\kappa_{3}{(\Delta{n})}\kappa_{2}{(N_{W})}\nonumber\\
&+&3\kappa_{2}^{2}{(\Delta{n})}\kappa_{2}{(N_{W})} +6\langle{\Delta{n}}\rangle^{2}\kappa_{2}{(\Delta{n})}\kappa_{3}{(N_{W})}+\langle{\Delta{n}}\rangle^4\kappa_{4}{(N_{W})},
\end{eqnarray}
where $\kappa_{n}{(\Delta{N})}$ and $\kappa_{n}{(\Delta{n})}$ are the measured cumulants and cumulants of net-quantities produced by each "source" which is assumed to be the number of participant ($N_{W}$) respectively.
\section{Analysis method}
\begin{figure}[htbp]   
\begin{center}   
\includegraphics[width=130mm]{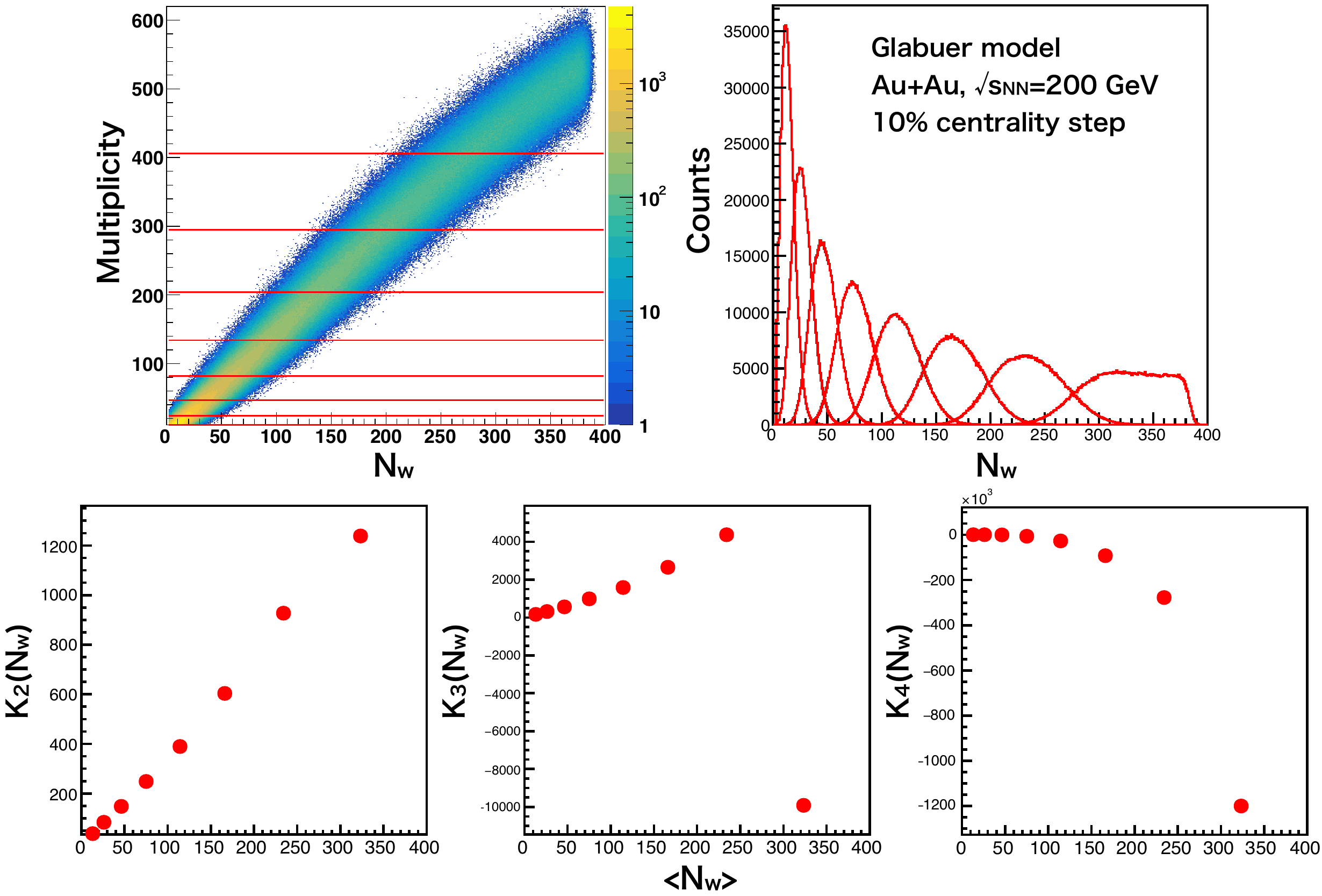}     
\end{center}
\caption{Correlation between multiplicity and $N_{W}$ by Glauber simulation (top left). $N_{W}$ distributions for each centrality (top right). Second to fourth-order cumulants as a function of $\langle{N_{W}}\rangle$ (bottom). Number of events are 500 Million.}      
\label{fig:pro_fig1} 
\end{figure}

%It is necessary to measure the cumulants of $N_{W}$ for VFC but $N_{W}$ can not be measured by experiment directly.
In order to estimate the VFC corrections, we need to determine the cumulants of $N_W$ distribution in the Eq.~(\ref{eq_vfc2})-(\ref{eq_vfc4}). Since we cannot directly measure the $N_W$ in experiment, one simple way is to use the Glauber model. 
In Glauber model, the final state multiplicity, which is used for centrality determination, is produced from each source independently.
This model is called Independent Particle Production (IPP) model because particles are produced from each source independently.
The number of source ($N_{source}$) is defined by two-component model as $N_{source}=(1-x)\frac{N_{W}}{2}+xN_{coll}$,
where $N_{coll}$ represents the number of collisions.
The number of produced particle from each source is fluctuating under the Negative Binomial Distribution (NBD),
and the parameters of Glauber and NBD are the same as net-charge analysis from STAR in Au+Au collisions at $\sqrt{s_{NN}}=200$ GeV \cite{pubnetc}.  
The top left panel of Fig.~\ref{fig:pro_fig1} shows the correlation between multiplicity and $N_{W}$, where red line represent the 10 \% step centralities divided by multiplicity.
The top right panel shows the $N_{W}$ distributions for each centrality, and second to fourth-order cumulants as a function of $N_{W}$ are shown in the bottom panels.
%In addition, two independent Poisson distributions are generated from each participant and then we reconstruct the cumulants.
Next, particles of interested, whose event-by-event distributions are analyzed, are generated from each participant nucleons independently (IPP) based on two Poisson distributions.
Cumulants of net-particle distribution should thus include VF defined by the Glauber model, which can subtracted by using Eq.~(\ref{eq_vfc1})-(\ref{eq_vfc4}).
The parameters of the Poisson distributions are determined that number of positively and negatively charged particles describe the real experiment respectively.
On the other hand, in UrQMD simulation, $N_{W}$ can be obtained directly.
Cumulants are also measured by using UrQMD approach in addition to toy model approach.

\section{Results}
\subsection{Toy model approach}
\begin{figure}[htbp]   
\begin{center}   
\includegraphics[width=155mm]{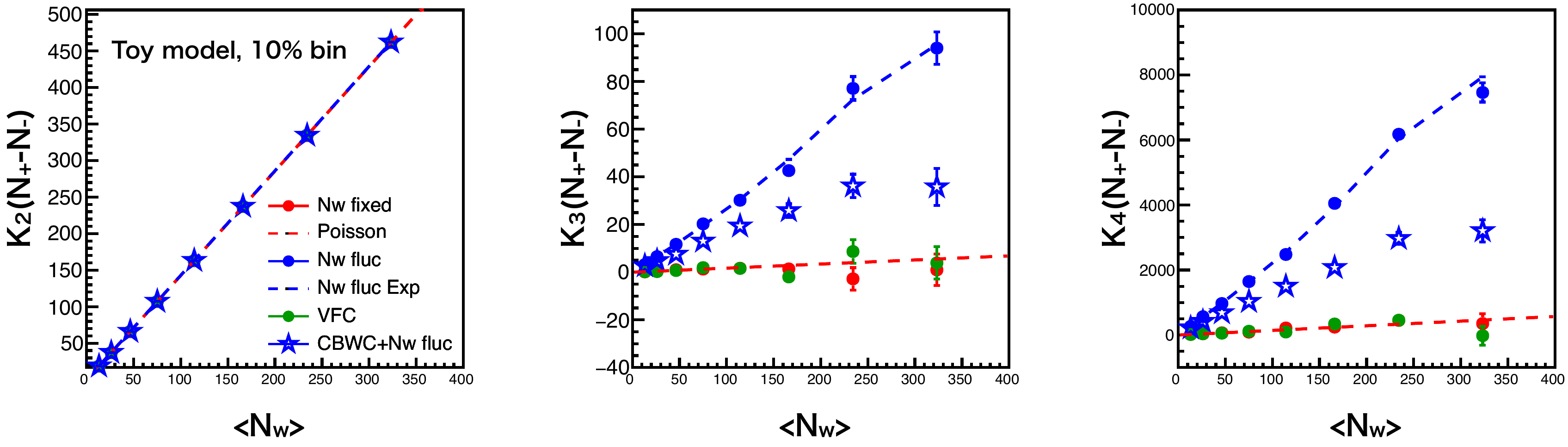}    
\end{center}
\caption{From second to fourth-order cumulants as a function of $\langle{N_{W}}\rangle$ by using toy model for for 10\% centrality step.}      
\label{fig:Toymodel} 
\end{figure}

Fig.~\ref{fig:Toymodel} shows the second to fourth-order cumulants of net-charge distribution as a function of $\langle{N_{W}}\rangle$ by using toy model for 10\% centrality step.
%Red symbols represents $N_{W}$ fixed results which means that $\langle{N_{W}}\rangle$ of each centrality was used instead of $N_{W}$.
%Therefore, red markers do not include the effect from VF.
For red points, $N_W$ is fixed at the value of the averaged number of participant nucleons ($\langle{N_{W}}\rangle$) in each centrality bin, they thus do not include VF.
Blue symbols include the fluctuation of the $N_{W}$ in each centrality.
Red and blue dotted line show the Poisson baseline and the expectation line of $N_{W}$ fluctuation which is estimated from Eq.~(\ref{eq_vfc1})-(\ref{eq_vfc4}) respectively.
$N_{W}$ fixed results (red) are consistent within Poisson baseline, and $N_{W}$ fluctuation results (blue) are also consistent with the baseline in all cases.
$K_{2}(N_{+}-N_{-})$ which corresponds to $K_{2}(\Delta{n})$ in Eq.~(\ref{eq_vfc2}) is not affected by VF. This is because small $\Delta{n}$ leads to small VF according to Eq.~(\ref{eq_vfc2}). 
For $K_{3}$ and $K_{4}$, $N_{W}$ fluctuation results are larger than $N_{W}$ fixed results which means that $N_{W}$ fluctuation results are
enhanced by VF.

\begin{figure}[htbp]   
\begin{center}   
\includegraphics[width=130mm]{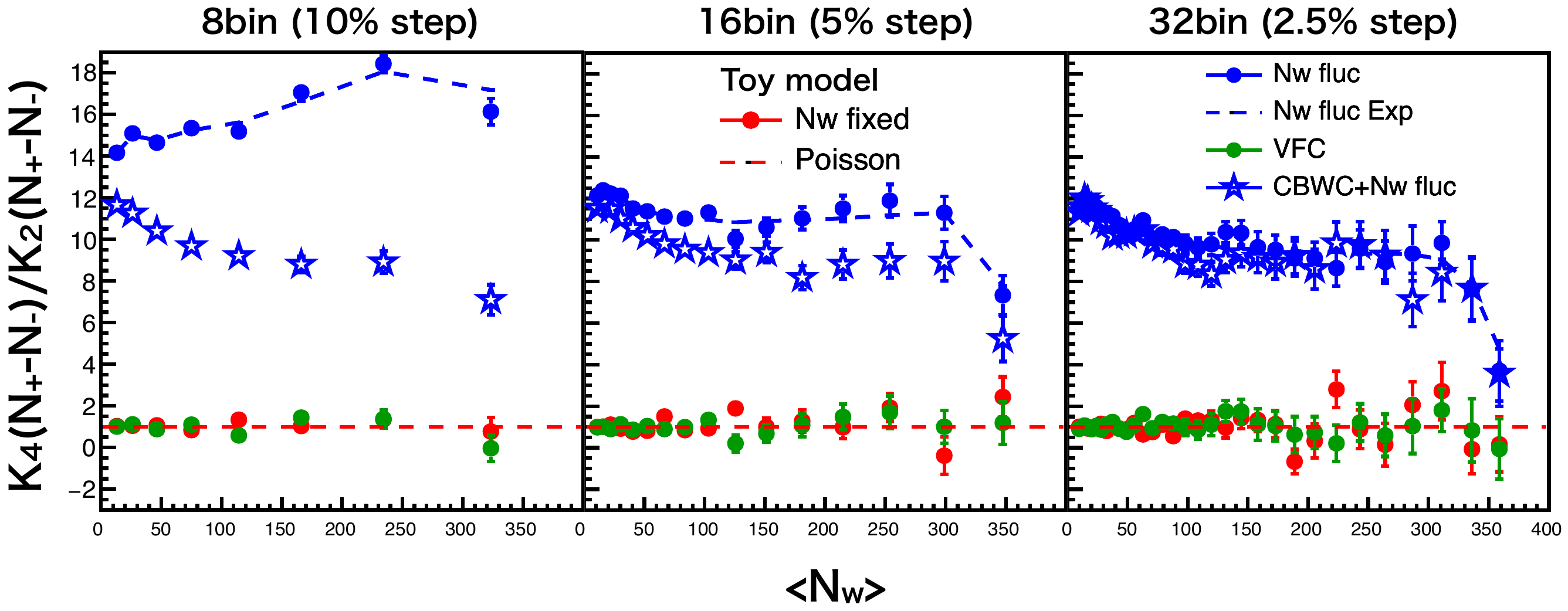}     
\end{center}
\caption{$\kappa\sigma^2$ as a function of mean number of participant by using toy model for 10\% (left), 5\% (middle) and 2.5\% (right) centrality step. The color and marker differences are the same as Fig.~\ref{fig:Toymodel}.}      
\label{fig:Toymodel_bin} 
\end{figure}

Then, we tried both CBWC and VFC to subtract VF from $N_W$ fluc results.
%The green and star symbols represent the two methods applied to subtract the effect of the fluctuation. Then you can compare that green describes the red points better, star symbols are subtracting some effect, but not all.
VFC results (green) are consistent with $N_{W}$ fixed results (red) which means that VFC works well in this model.
On the other hand, CBWC results are smaller than $N_{W}$ fluctuation results but larger than $N_{W}$ fixed results.
This results mean that CBWC can reduce VF but can not completely eliminate the VF.
Therefore, in toy model case, CBWC is not enough and VFC works well.
Fig.~\ref{fig:Toymodel_bin} shows the $\kappa\sigma^2$ of net-charge distribution as a function of $\langle{N_{W}}\rangle$ for 10\%, 5\% and 2.5\% 
centrality step.
%In 10\% centrality step, CBWC results are smaller than $N_{W}$ fluctuation results which trends are same as Fig.~\ref{fig:Toymodel}.
In 10\% centrality step, CBWC results contain larger VF compared to the results with 5\% and 2.5\% step centrality.
However, the differences between CBWC and $N_{W}$ fluctuation results become smaller in 5\% centrality step and consistent in
2.5\% step.
This results imply that 2.5\% centrality step can reduce VF as well as CBWC.
However, there remain VF in both CBWC and $N_{W}$ fluctuation results in any case.
\subsection{UrQMD approach}
\begin{figure}[htbp]   
\begin{center}   
\includegraphics[width=155mm]{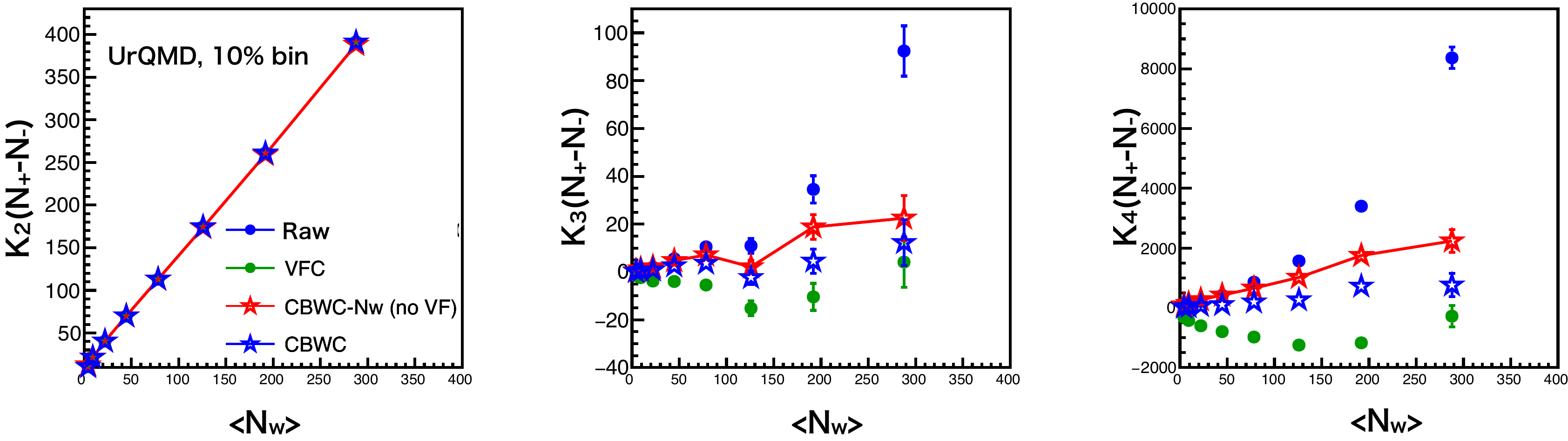}     
\end{center}
\caption{From second to fourth-order cumulants as a function of $\langle{N_{W}}\rangle$ by using UrQMD model simulation for 10\% centrality step. The color and marker differences are the same as Fig.~\ref{fig:Toymodel}.}      
\label{fig:VFC_UrQMD} 
\end{figure}

Fig.~\ref{fig:VFC_UrQMD} shows the second to fourth-order cumulants of net-charge distribution as a function of $\langle{N_{W}}\rangle$ by using UrQMD model for 10\% centrality step. Red open star symbols "CBWC-$N_{W}$" mean that CBWC is applied for each $N_{W}$ bin.
Standard CBWC is applied for each multiplicity bin which is represented by blue open star symbol.
CBWC-$N_{W}$ results are considered as "no-VF" results which correspond to the red round symbol in the toy model case.
%Blue symbols represent $N_{W}$ fluctuating, which mean raw and no correction, results and green symbols are VFC results.
Blue symbols contain VF without any corrections, and VFC results are shown in green markers.
As discussed in previous section, $K_{2}$ is not affected by VF due to the small value of $\Delta{n}$.
However, trends at $K_{3}$ and $K_{4}$ are not consistent with toy model case.
For example, CBWC results are smaller than CBWC-$N_{W}$ results, and VFC results are smaller than both of them.
VFC seems over correction and does not work well.
One of the reason could be that IPP is broken in UrQMD model.
\section{Conclusions}
Importance of the volume fluctuation correction on higher cumulants are presented by using toy model assuming IPP and UrQMD simulation.
From these studies, 2.5\% centrality division can reduce VF as well as CBWC but 5\% and 10\% centrality divisions include the effect from VF.
%From these studies, 2.5\% centrality division can reduce VF as well as CBWC.
In toy model, even though CBWC has applied, effect from VF can not be removed completely and VFC works well.
However, VFC does not work well in UrQMD model, which could be because IPP model is broken in UrQMD.
Therefore, we have to consider these effect if VFC is applied to experimental data, and further studies are needed in order to fully understand 
how to correctly subtract the VF from the measured cumulants.

\end{document}